\documentclass[aps,prl,twocolumn,superscriptaddress,showpacs]{revtex4-1}
\pdfoutput=1
\usepackage{graphicx}
\usepackage{color}
\usepackage{placeins}

\begin{document}

\title{Multi-step approach to microscopic models for frustrated quantum magnets
\\ --   the case of the natural mineral azurite}

\author{Harald Jeschke}
\affiliation{Institut f\"ur The\-o\-re\-tische Physik, Goethe-Uni\-ver\-si\-t\"at Frank\-furt am Main, 60438 Frankfurt am Main, Germany}
\author{Ingo Opahle}
\affiliation{Institut f\"ur The\-o\-re\-tische Physik, Goethe-Uni\-ver\-si\-t\"at Frank\-furt am Main, 60438 Frankfurt am Main, Germany}
\author{Hem Kandpal}
\affiliation{IFW Dresden, POB 270116, 01171 Dresden, Germany}
\author{Roser Valent\'{\i}}
\affiliation{Institut f\"ur The\-o\-re\-tische Physik, Goethe-Uni\-ver\-si\-t\"at Frank\-furt am Main, 60438 Frankfurt am Main, Germany}

\author{Hena Das}
\affiliation{Satyandranath Bose National Centre for Basic Sciences, Kolkata 700098, India}
\author{Tanusri Saha-Dasgupta}
\affiliation{Satyandranath Bose National Centre for Basic Sciences, Kolkata 700098, India}

\author{Oleg Janson}
\affiliation{Max Planck Institute for Chemical Physics of Solids, 01187 Dresden, Germany}
\author{Helge Rosner}
\affiliation{Max Planck Institute for Chemical Physics of Solids, 01187 Dresden, Germany}

\author{Andreas Br\"uhl}
\affiliation{Physikalisches Institut, Goethe-Uni\-ver\-si\-t\"at Frank\-furt am Main, 60438 Frankfurt am Main, Germany}
\author{Bernd Wolf}
\affiliation{Physikalisches Institut, Goethe-Uni\-ver\-si\-t\"at Frank\-furt am Main, 60438 Frankfurt am Main, Germany}
\author{Michael Lang}
\affiliation{Physikalisches Institut, Goethe-Uni\-ver\-si\-t\"at Frank\-furt am Main, 60438 Frankfurt am Main, Germany}

\author{Johannes Richter}
\affiliation{Institut f\"{u}r Theoretische Physik, Universit\"{a}t Magdeburg, P.O. Box 4120, 39016 Magdeburg, Germany}

\author{Shijie Hu}
\affiliation{Department of Physics, Renmin University of China, Beijing 100872, China}
\author{Xiaoqun Wang}
\affiliation{Department of Physics, Renmin University of China, Beijing 100872, China}

\author{Robert Peters}
\affiliation{Department of Physics, Graduate School of Science, Kyoto University, Kyoto 606-8502, Japan}

\author{Thomas Pruschke}
\affiliation{Institut f\"ur Theoretische Physik, Georg-August-Universit\"at G\"ottingen, 37077 G\"ottingen, Germany}
\author{Andreas Honecker}
\affiliation{Institut f\"ur Theoretische Physik, Georg-August-Universit\"at G\"ottingen, 37077 G\"ottingen, Germany}

\begin{abstract}

The natural mineral azurite Cu$_3$(CO$_3$)$_2$(OH)$_2$ is a
frustrated magnet displaying unusual and controversially discussed
magnetic behavior. Motivated by the lack of a unified description
for this system, we perform a theoretical study based on density
functional theory as well as state-of-the-art numerical many-body
calculations. We propose an effective generalized spin-1/2 diamond
chain model which provides a consistent description of experiments:
low-temperature magnetization, inelastic neutron scattering, nuclear
magnetic resonance measurements, magnetic susceptibility as well as
new specific heat measurements. With this study we demonstrate that
the balanced combination of first principles with powerful many-body
methods successfully describes the behavior of this frustrated
material.

\end{abstract}

\date{December 5, 2010; revised April 21, 2011}


\maketitle

\begin{figure*}[t]
\includegraphics[width=0.95\textwidth]{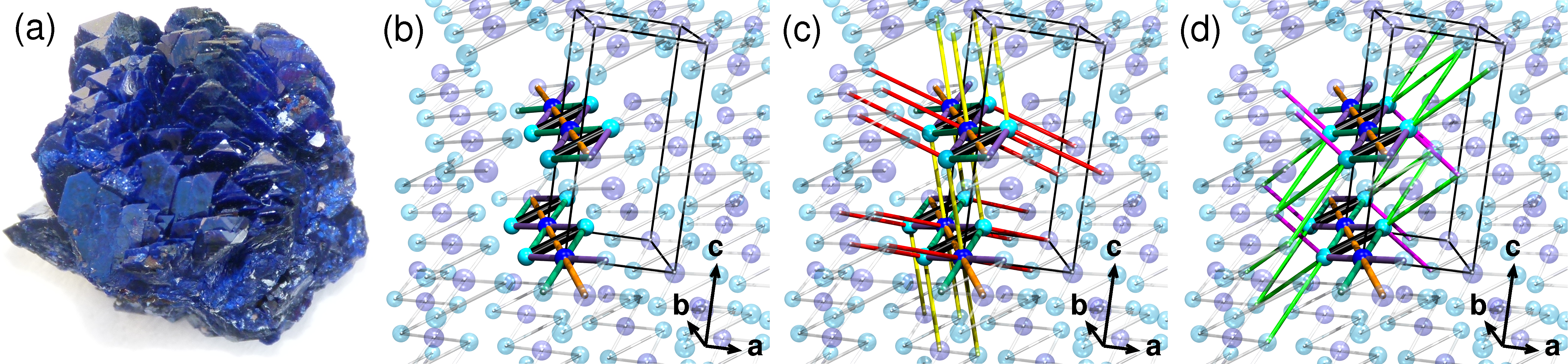}
\caption{(a) Example of an azurite crystal aggregate. (b)-(d) Arrangement of Cu$^{2+}$ ions in the structure of
    azurite. The two inequivalent Cu$^{2+}$ ions form dimers (cyan)
    and monomers (blue).
    (b) Most important exchange
    paths within the diamond chain running along the
     $b$-axis: Dimer coupling $J_2$ (black),
    dimer-monomer couplings $J_1$ and $J_3$ (magenta and green), and
    monomer-monomer coupling $J_{\text{m}}$ (orange).  (c)-(d)
    Three-dimensional couplings between diamond chains, connecting
    (c) monomer and dimer ions: $J_5$ (yellow) and $J_6$ (red) and  (d)
    dimer ions only: $J_4$ (pink) and $J_7$ (light green).  }
\label{FIG:structure}
\end{figure*}

The natural mineral azurite Cu$_3$(CO$_3$)$_2$(OH)$_2$ has been used
as a blue pigment since the time of the ancient Egyptians; the
beautiful intense blue color (see Fig.~\ref{FIG:structure}~(a)) is due
to the crystal field splitting of Cu $3d$ orbitals in square planar
coordination.  More recently, the discovery of a plateau at 1/3 of the
saturation value in the low-temperature magnetization
curve~\cite{KikuchiA,KikuchiE} has triggered intensive interest in the
magnetic properties of azurite.
{}From the point of view of magnetism, the most important structural
motives~\cite{Zigan:72} are diamond-like chains which are formed by
the spin-1/2 copper atoms (Fig.~\ref{FIG:structure}~(b)). If all
exchange constants were antiferromagnetic, azurite would fall into the
class of geometrically
frustrated magnets. These systems are fascinating since the competition
of the magnetic interactions suppresses classically ordered states and
may give rise to new states of matter with exotic excitations (see
Ref.~\cite{HFMbook} for a recent review).  In particular, for a
certain class of frustrated magnets including diamond chains, one
expects localized (dispersionless) many-body states at high magnetic
fields~\cite{LTP2007};
indeed inelastic neutron scattering (INS) on azurite exhibits
an almost dispersionless branch of excitations~\cite{azuritINS}.

\begin{figure}[t]
\includegraphics[width=\columnwidth]{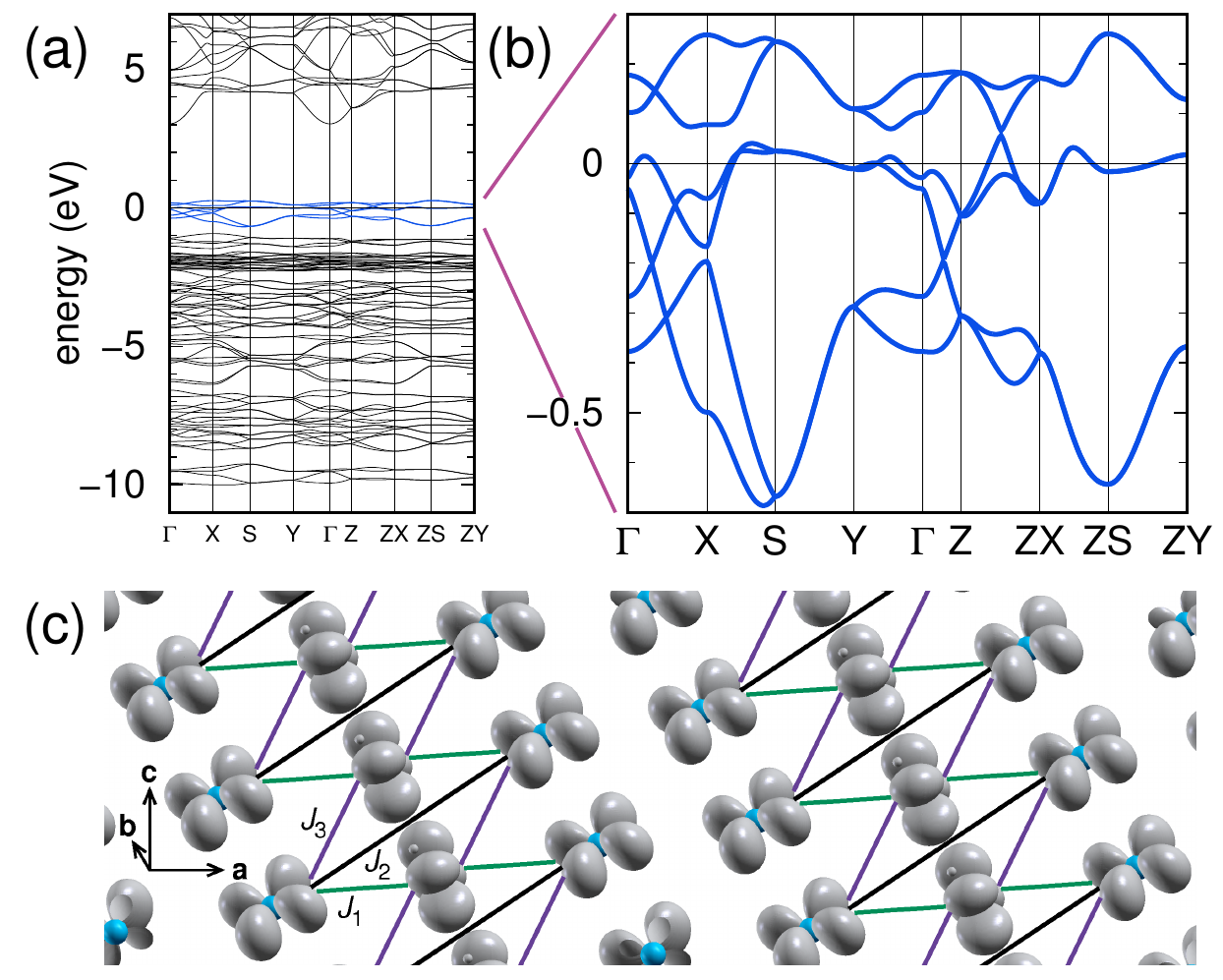}
\caption{Electronic structure of azurite, calculated with
    FPLAPW. (a) Band structure in a wide energy window.
 At the   Fermi level the bands are dominantly of Cu $3d_{x^2-y^2}$ character
    (blue bands).  (b) Blow-up of the six bands at the Fermi
    level.  (c)
    Electron density above $E=-0.75$~eV for an isovalue of 0.1~e/a.u.$^3$.
     All density is centered at
    the Cu sites and it has $3d_{x^2-y^2}$ symmetry.  }
\label{FIG:bandstructure}
\end{figure}

\begin{figure}[t]
\includegraphics[width=\columnwidth]{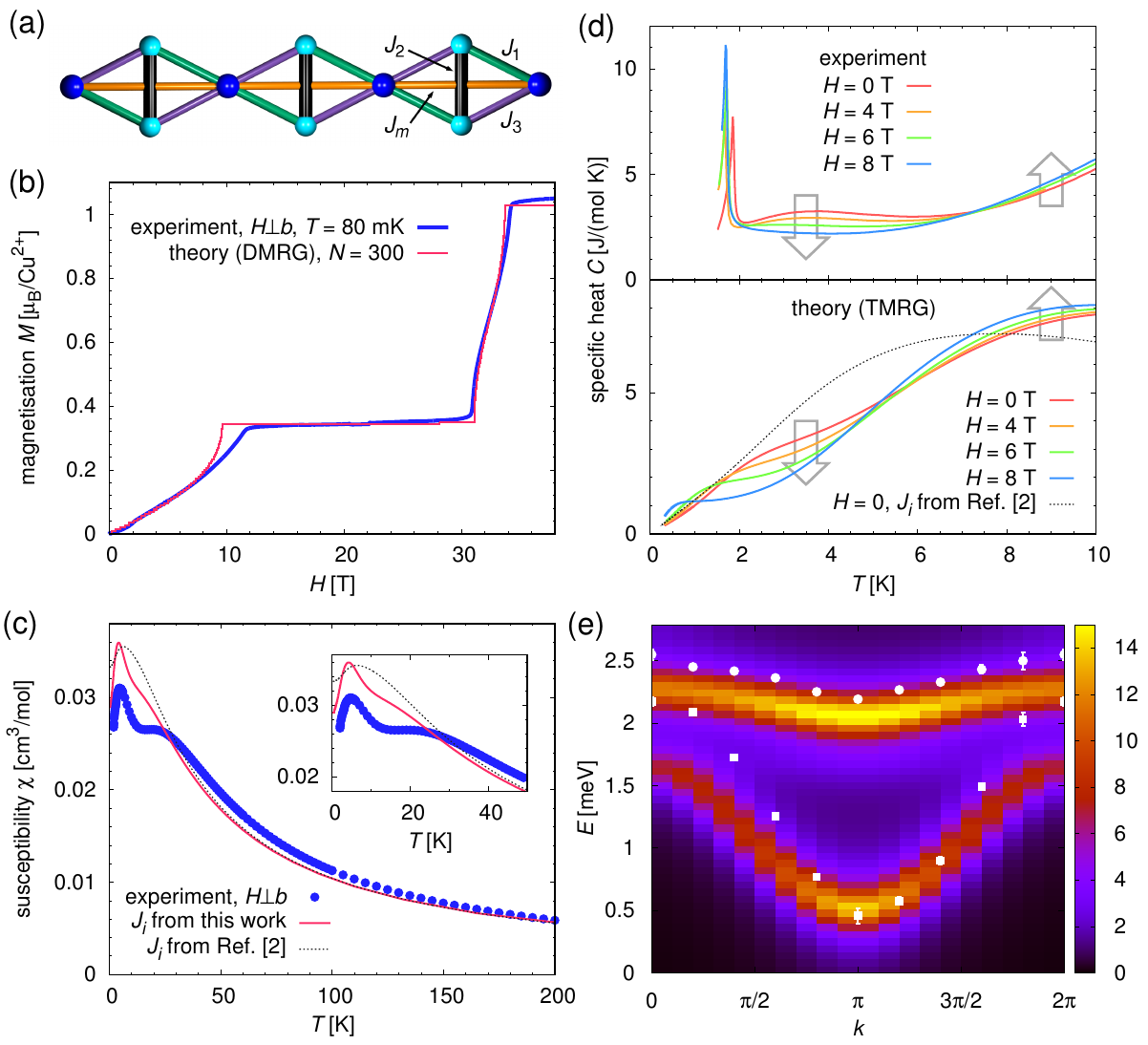}
\caption{
    (a) Generalized diamond chain model.  (b) Comparison of
    computations for the magnetization curve for $T=0$ and $N=300$
    spins with experimental data at $T=80$~mK for $H \perp
    b$~\cite{KikuchiE}.  (c) Experimental and theoretical
    zero-field magnetic susceptibility.  (d) Experimental (upper
    panel) and theoretical (lower panel)  specific
    heat results for various $H$ fields. 
 Arrows indicate the response to increasing magnetic
    field.
    (e) Theoretical transverse dynamic structure factor on the
    1/3 plateau ($H\approx 14$~T) and peak positions of
    INS spectra from Ref.~\cite{azuritINS} (white symbols).
    Color-coding represents the intensity in arbitrary units. }
\label{fig:comparison}
\end{figure}

There have been a number of
attempts~\cite{KikuchiA,GuSu1,GuSu2,azuritINS,MiLu,Whangbo} to
derive a microscopic model for the complex
magnetic properties of azurite. The results are, however, contradictory
and up to now none of these models was able to
yield a fully consistent picture of the experimentally observed
behavior. Some authors favor a diamond chain model with
all exchanges antiferromagnetic~\cite{KikuchiA,MiLu,KikuchiB} while other
authors proposed one of the dominant exchange constants
to be ferromagnetic~\cite{GuSu1,GuSu2,azuritINS}. 
Even more, Ref.~\cite{Whangbo} has argued
that interchain coupling is important in azurite. The latter may be in
agreement with the observation of a magnetic ordering transition
at about $2$~K~\cite{KikuchiA,Forstat,gibson10,rule10}, but raises the question why
no dispersion perpendicular
to the chain direction is observed by INS \cite{azuritINS}.

In the present work, we combine first principles
density functional theory (DFT) calculations with model computations
based on different variants of the density-matrix renormalization
group (DMRG) method~\cite{DMRGa,TMRG,dDMRG} (see also Section \ref{sec:DMRG})
and resolve the underlying model for azurite. We find that an effective
generalized spin-1/2 diamond chain model with a dominant
next-nearest-neighbor antiferromagnetic Cu  dimer
coupling $J_2$, two antiferromagnetic nearest- and
third-nearest-neighbor Cu dimer-monomer exchanges $J_1$ and $J_3$,
and a significant
direct Cu monomer-monomer exchange $J_{\text{m}}$ (see
Figs.~\ref{FIG:structure} (b) and \ref{fig:comparison}~(a))
explains a broad range of
experiments on azurite~\cite{KikuchiA,KikuchiE,azuritINS,azuritNMR}
and resolves the existing controversies.

Since the experimentally determined positions of the lighter atoms in
a structure usually carry larger error bars than those of more heavy
elements, we first performed a Car-Parrinello molecular dynamics
calculation~\cite{Parrinello,Bloechl} in order to optimize the
positions of the C, O, and H atoms in azurite.  With the optimized structure with a total of 30 atoms in the unit cell
we  determined the electronic properties of
azurite~\cite{WIEN}.  The  band structure shows six narrow
Cu 3$d_{x^2-y^2}$ bands at the Fermi level -- corresponding to the six
Cu atoms per unit cell -- separated by an energy of $0.9$~eV from the
occupied Cu 3$d_{z^2}$ bands and by a gap of $E_g\simeq 3$~eV from the
higher unoccupied bands (Fig.~\ref{FIG:bandstructure}~(a) and 
(b)).  Fig.~\ref{FIG:bandstructure}~(c) shows a charge density
isosurface where a $d_{x^2-y^2}$ symmetry of the Cu $d$ orbitals is
evident without contribution of $d_{z^2}$ character, contrary to
previous suggestions~\cite{Belokoneva,comment}. 

While the GGA calculation describes this system as metallic, the
insulating behavior is correctly given within the GGA+U approach (see
below). Here we first analyze the interaction paths based on the GGA
bandstructure.  We perform $N$-th order muffin tin orbital (NMTO)
downfolding~\cite{NMTO,valenti02} to obtain the tight-binding
Hamiltonian parameters $t_i$ describing the six Cu 3$d_{x^2-y^2}$
bands (see Fig.~\ref{FIG:bandstructure}~(b)). Under
the assumption that the exchange couplings are antiferromagnetic,
we can estimate the magnitude of the exchange couplings via second-order
perturbation theory: $J_i^{\text{AFM}}=4t_i^2/U$ where $U$ is the Cu
$3d$ onsite Coulomb interaction strength.  From this
analysis we identify six further relevant couplings in addition to
$J_1$, $J_2$ and $J_3$: the monomer-monomer exchange $J_{\text{m}}$ also
considered by Rule {\em et al.}~\cite{azuritINS} and 
the nearest-neighbor Cu dimer interaction $J_{\text{d}}$
along the chain. In addition, $J_4$ and $J_7$ provide couplings
between Cu dimer atoms in neighboring chains, whereas $J_5$ and $J_6$
correspond to Cu monomer-dimer interchain-interactions.
The interaction paths between  chains are visualized in
Fig.~\ref{FIG:structure} (c) and (d).

\begin{table}[t]
\begin{tabular}{l|c|c|c|c|c|c|c|c|c}
 & $J_1$ & $J_2$ & $J_3$ & $J_4$ & $J_5$ & $J_6$ & $J_7$ &$J_{\text{m}}$ & $J_{\text{d}}$\\\hline\hline
{\bf 1} full model & 13.5 & 42.8 & 12.5 & 2.7 & 0.6 & 4.4 & --1.7 & 2.6 & --0.4 \\\hline
{\bf 2} minimal model & 17.9 & 43.9 & 12.0 &--&--&--&--& 2.4 &--\\\hline
{\bf 3} refined model& 15.51 & 33 & 6.93  &--&--&--&--& 4.62 &--\\\hline\hline
{\bf 4} Ref.~\cite{KikuchiA} model & 19 & 24 & 8.6 &--&--&--&--&--&--
\end{tabular}
\caption{Exchange constants in Kelvin (K) derived from FPLO
    GGA+U calculations with $U=8$~eV and $J_{\protect\text{H}}=1$~eV
    for the various model steps considered in the present work 
     (see text for explanation).
    The error margin for each $J_i$ in the third line is estimated to
    be of the order 1 to 2~K.}
\label{TAB:Ji}
\end{table}

Next, we obtain the correct sign (ferro- or antiferromagnetic) and magnitude of the
$J_i$  from total energy
calculations for different Cu spin configurations in supercells with
up to 60 atoms. We employ the full potential local orbital (FPLO)
method~\cite{FPLO} with the GGA+U functional for $U = 4$, $6$, and $8$~eV.  
We map the energy differences of the frozen collinear spin
configurations onto a spin-1/2 Heisenberg model and evaluate the
exchange constants $J$ in a dimer approximation~\cite{ingo06}.  The
nine relevant Cu-Cu  interaction paths obtained from the downfolding
calculations have been probed with 10 different
antiferromagnetic spin configurations together with 
the ferromagnetic configuration.
The result for a choice of $U= 8$~eV and $J_{\text{H}} = 1$~eV is
shown in the first line of Table~\ref{TAB:Ji}.  As expected from
experimental observations, $J_2$ dominates and exhibits a $1/U$
dependence (see Section \ref{sec:DFT}). The two couplings $J_1$ and $J_3$ are very
similar in magnitude, suggesting an almost symmetric diamond chain.
We observe that except for $J_1$, $J_2$, and $J_3$, the coupling
strengths are of the order of a few Kelvin.  Comparing our
set of parameters in Table~\ref{TAB:Ji}, line {\bf 1} to that obtained
in Ref.~\cite{Whangbo}, the main differences are that we determined
the additional 3D couplings $J_4$, $J_5$, and $J_7$, and our value for
$J_{\text{m}}$, double-checked with two full potential
methods~\cite{WIEN,FPLO}, is clearly nonzero.

At first sight, the fact that interchain coupling turns out to be
appreciable is surprising because INS
did not observe any dispersion perpendicular to the chain
direction~\cite{azuritINS}.  However, since the dimer exchange $J_2$
dominates, one can use perturbative arguments along the lines of
Ref.~\cite{HoL} to show that there are no low-energy excitations
dispersing perpendicular to the chains. The essential ingredients of
the argument are that (i) the interchain exchange constants $J_4$ to
$J_7$ are small compared to $J_2$ and (ii) they connect only to dimers
of the neighboring chains (compare Figs.~\ref{FIG:structure}~(c)
and (d)), {\it i.e.}, they contribute only in second or third
order in perturbation theory (see Section \ref{sec:pert});
this would suggest using an effective one-dimensional model with the
values of $J_1$, $J_2$, $J_3$, and $J_{\text{m}}$ adjusted to
incorporate the effect of the three-dimensional couplings.
Table~\ref{TAB:Ji} line {\bf 2} shows the results obtained by solving
the 10 spin configurations only for the diamond chain couplings. This
corresponds to an averaging over the 3D couplings and translates into
a significant asymmetry of the diamond chain $J_1 > J_3$.  The
effective one-dimensional model has the additional advantage that it is
amenable to detailed quantum mechanical model calculations, thus
allowing a quantitative comparison with experimental data for azurite.

From these results salient experimental features of azurite can
already be understood at a qualitative level: two thirds of the
Cu$^{2+}$ spins are strongly bound by $J_2$ into dimer singlets while
another third consists of monomer spins which interact weakly by
$J_{\text{m}}$ and additional effective monomer-monomer interactions
which are generated by integrating out the dimers.  In an applied
magnetic field, the monomer spins are therefore polarized first while
the dimer spins remain in the singlet state, giving rise to the 1/3
plateau~\cite{KikuchiA,KikuchiE,azuritNMR}.  Furthermore, the two energy
scales, {\it i.e.}, the low-energy scale given by the monomer-monomer
interactions and the high-energy scale associated to the dimers give
rise to the double-peak structures observed in the magnetic
susceptibility~\cite{KikuchiA} and the specific heat
\cite{KikuchiA,azuritINS}. Finally, we expect a band of low-energy
monomer excitations dispersing along the chain direction and a band of
dimer excitations at higher energies whose dispersion is additionally
suppressed by the competition of $J_1$ and $J_3$, as indeed observed
by INS~\cite{azuritINS}.

We will now show that we can also describe these experimental results
quantitatively. The DFT results leave some freedom concerning the
overall energy scale, however the ratios of the $J_i$ are
expected to be subject only to small errors~\cite{kandpal09}.  We
therefore first slightly refined the parameter ratios using the
magnetization and INS experiments, leading to
${J_1}/{J_2} = 0.47$, ${J_3}/{J_2} = 0.21$, and ${J_{\text{m}}}/{J_2}
= 0.14$.  The global energy scale is finally adjusted to the
magnetization curve (see below) and we obtain the exchange coupling
constants $J_i$ in Table~\ref{TAB:Ji}, line {\bf 3}.

In order to fully account for the quantum nature of the spins residing
on the Cu$^{2+}$ ions, we use a spin 1/2 Heisenberg model ${\cal H} =
\sum_{\langle i,j \rangle} J_{i,j} \vec{S}_i \cdot \vec{S}_j -
g\,\mu_B\,H\,\sum_i S_i^z$, where $\vec{S}_i$ are spin 1/2 operators,
$J_{i,j}$ is the exchange constant connecting sites $i$ and $j$ (see
Fig.~\ref{fig:comparison}~(a)), $H$ an external magnetic field and
$\mu_B$ the Bohr magneton.  The gyromagnetic ratio
$g$ is set to $2.06$ \cite{Ohta}.

Fig.~\ref{fig:comparison}~(b) shows a comparison for the experimental
and computed magnetization curves. The
overall energy scale is $J_2=33$~K, leading to our final parameter set
in Table~\ref{TAB:Ji}, line {\bf 3}. The agreement of the theoretical
magnetization curve in Fig.~\ref{fig:comparison}~(b) with the
experimental result for $H \perp b$~\cite{KikuchiE} is excellent.
Note that the experimental curve for $H \perp b$ exhibits a nice
plateau as expected for a Heisenberg model whereas for $H
\parallel b$ the plateau is washed out~\cite{KikuchiA}, indicative of
non-commuting fields. Therefore we compare our results for the
isotropic Heisenberg model with experiments for $H \perp b$.
We find that dimer spins should be about 2.7{\%} polarized each
(see Section \ref{sec:strucM1o3}),
{\it i.e.}, dimers are essentially in the singlet state whereas the
single ``monomer'' spins are almost fully polarized in the 1/3
plateau. This is qualitatively consistent with recent $^{63,65}$Cu
NMR~\cite{azuritNMR}.

At this stage, the values of all $J_i$ are fixed and we have a
parameter-free prediction of the magnetic susceptibility
$\chi$. Fig.~\ref{fig:comparison}~(c) compares our computations \cite{TMRG}
with our measurement of the magnetic susceptibility of azurite for
$H \perp b$, which is
very similar to the original experiment of Ref.~\cite{KikuchiA}.  Our
parameter set (Table~\ref{TAB:Ji}, line {\bf 3}) leads to a small, but
qualitative improvement compared to the original one
of~\cite{KikuchiA} (see Table~\ref{TAB:Ji}, line {\bf 4}): we
reproduce a double-peak-like structure at the correct temperatures
whereas only a single peak~\cite{GuSu1} is obtained with the parameters of~\cite{KikuchiA}.

Analogous to the magnetic susceptibility, we also have a
parameter-free prediction for the magnetic specific heat. At zero
field, two anomalies have been observed in the specific heat at
$T=18$~K~\cite{KikuchiA} and
$T=4$~K~\cite{KikuchiA,azuritINS}. Fig.~\ref{fig:comparison}~(d)
compares the field-dependence of the experimental specific heat with
results calculated \cite{TMRG}
for the parameters of Table~\ref{TAB:Ji}, line {\bf
  3}. The sharp peak in the experimental curves slightly below
$2$~K~\cite{KikuchiA,Forstat} signals an ordering transition which is
out of reach of a one-dimensional model.  Nevertheless, not only are
the numerical values of the specific heat for $2\:\text{K} < T
\lesssim 10$~K comparable between theory and experiment, but also
important features are reproduced correctly: (i) a low-temperature
peak appears for $H=0$ at $T\approx3$ to $4$~K. Note that this
low-temperature peak at $H=0$ is absent~\cite{GuSu2} for the original
parameter set of Ref.~\cite{KikuchiA} (compare the dashed curve in the
lower panel of Fig.~\ref{fig:comparison}~(d)). (ii) The
low-temperature peak is gradually suppressed by an applied field, as
emphasized by down arrows in the figure. (iii) In the temperature
range $7\:\text{K} \le T \le 10\:\text{K}$, the specific heat
increases not only with temperature but also with increasing magnetic
field (marked by up arrows).

Fig.~\ref{fig:comparison}~(e) shows our numerical result \cite{dDMRG}
(see also Section \ref{sec:DMRG})
for the transverse dynamic structure factor on the 1/3 plateau as a function
of momentum transfer $k$ along the chain direction and energy $E$.
The peak values of the dynamic structure factor trace two dispersion
curves nicely. Comparison with the corresponding
INS results~\cite{azuritINS} (%
white symbols in Fig.~\ref{fig:comparison}~(e)) shows that
the computed ratio of the bandwidths is extremely close to
the experimental value of about $1/6$ \cite{footnote}.
Also  the total intensities in the peaks compare favorably with the
experimental results~\cite{azuritINS}.

To summarize, we have shown that the combination of first principles
DFT with state-of-the-art many-body calculations successfully provides a
microscopic model for the frustrated magnet azurite, which
explains a wide range of
experiments~\cite{KikuchiA,KikuchiE,azuritINS,azuritNMR}. We believe
that attempts to fit such a range of experiments, using at least four
exchange constants $J_i$, are bound to fail. Hence, the guiding DFT
computations were essential.
There are several issues for further experimental and theoretical
study (see Section \ref{sec:perspectives}). In particular, the implications of the full
three-dimensional model which we have derived remain to be explored.

\begin{acknowledgments}

We would like to thank H.\ Kikuchi and S.\ S\"ullow
for providing us with the
experimental data shown in Fig.~\ref{fig:comparison}~(b) and
(e), respectively. Useful
discussions with C.\ Berthier, M.\ Horvati\'c, and S.\ S\"ullow are
gratefully acknowledged. This work has been supported by the DFG
(SFB/TR~49, SFB~602, HO~2325/4-2, PR~298/10, and RI~615/16-1),
by the Helmholtz Association through HA216/EMMI,
the National Natural Science Foundation of China (NSFC),
and the JSPS together with the Humboldt Foundation for R.P.

\end{acknowledgments}

\setcounter{secnumdepth}{3}

\setcounter{table}{0}
\setcounter{figure}{0}

\renewcommand{\thetable}{S\Roman{table}}
\renewcommand{\thefigure}{S\arabic{figure}}
\renewcommand{\thesection}{}
\renewcommand{\thesubsection}{S\arabic{subsection}}

\section*{Supplementary material}

\subsection{Density functional calculations}

\label{sec:DFT}

Structure relaxations were performed with the Car-Parrinello projector
augmented wave (CP-PAW) method~\cite{Bloechl}. We employed a plane
wave cutoff of 30~Ryd for the plane wave part and of 120~Ryd for the
density, respectively, and we used the following sets of (s,p,d)
projector functions per angular momentum: Cu(2,2,2), O(2,2,1),
C(2,2,1) and H(2,0,0). We employed a $(4\times 4\times 4)$ k mesh and
the $P\,21/c$ symmetry was preserved during the relaxation with the
help of 60 constraints. The relaxation with
the generalized gradient approximation (GGA) functional~\cite{GGA}
resulted in small bond length changes of up to 4{\%} and in angle
changes up to 2$^\circ$ compared to the experimental data from
Ref.~\cite{Zigan:72}.

The relative strengths of the exchange pathways in azurite have been
obtained by using the electronic structure technique of muffin-tin
orbital (MTO) based {\it N}MTO-downfolding~\cite{NMTO,Andersen03}.

DFT calculations were performed with the full potential local orbital
method~\cite{FPLO} (FPLO), version 8.50, and the full potential
augmented plane wave (FLAPW) method as implemented in the
WIEN2k~\cite{WIEN} code, which has been used to crosscheck the FPLO
results for selected supercells. Total energies for different spin
configurations were obtained in the GGA+U formalism, employing both
the atomic limit (AL) as well as the around mean field (AMF) double
counting correction. The AL double counting correction turned out to
be the better choice for the calculation of a realistic set of model
parameters for azurite because the ratios $J_i/J_2$ are strongly
dependent on $U$ in the case of the AMF double counting
correction and can even adopt unphysical values ($J_1, \,J_3 > J_2$).

Table~\ref{TAB:JsALK} shows the complete set of exchange coupling
parameters $J_i$, obtained with the FPLO code (version 8.50) employing
the GGA+U functional with atomic limit double counting
correction. The calculations were repeated for three choices of the
Coulomb correlation strength, $U=4$~eV, $6$~eV and $8$~eV. The Hund's
rule coupling $J_{\text{H}}$ was chosen as $J_{\text{H}}=1$~eV. The dominant
coupling exhibits a proportionality to $1/U$. Note that
the relative importance of the monomer-monomer coupling increases as
$U$ is increased.

\begin{table}
\begin{tabular}{c||c|c|c|c|c|c|c|c|c}
$U$ [eV] & $J_1$ & $J_2$ & $J_3$ & $J_4$ & $J_5$ & $J_6$ & $J_7$ &
$J_{\text{m}}$ & $J_{\text{d}}$ \\\hline
4 & 34.1 & 145.4 & 35.4 & 5.9 & 2.9 & 16.2 & --1.7 & 5.9 & --1.7 \\
6 & 21.3 & 82.8 & 21.2 & 3.8 & 1.5 & 8.6 & --1.8 & 3.9 & --0.8 \\
8 & 13.5 & 42.8 & 12.5 & 2.7 & 0.6 & 4.4 & --1.7 & 2.6 & --0.4 \\
\end{tabular}
\caption{{Exchange constants in K derived from FPLO GGA+U calculations
    with the atomic limit double counting correction. Slater parameters 
    are chosen as $F_0=U$, $F_2=8.6$~eV and $F_4=5.4$~eV, {\it i.e.},
    $J_{\text{H}}=(F_2+F_4)/14=1$~eV.}}
\label{TAB:JsALK}
\end{table}

$J_2$ is antiferromagnetic and the dominant interaction.  One can
therefore apply perturbative considerations in $J_i/J_2$ and argue
that interchain excitations can be neglected to a first approximation
(see Section~\ref{sec:pert}).  The essential items are that the
interchain exchange constants $J_4$ to $J_7$ are small compared to
$J_2$ and that they connect only to dimers of the neighboring chains
(compare Figs.~1~(c) and (d)).

This suggests to reduce the set of interaction parameters to a minimal
model including $J_1$, $J_2$, $J_3$, and $J_{\text{m}}$ only. In order
to determine the effective values of these exchange constants
quantitatively, we performed a least-square fit for the energy
differences including only $J_1$, $J_2$, $J_3$, and $J_{\text{m}}$ in
the model and set all other parameters to zero. It should be noted
that the exchange constants obtained in this way are effective
parameters, which contain the effect of the remaining parameters not
included in the model as statistical average. Furthermore, it is also
important that the procedure of statistical averaging is done over a
sufficiently large manifold, since otherwise the effective parameters
are to some degree arbitrary.  The results are shown in
Table~\ref{TAB:JeffALK}.  It can be seen that in the minimal model the
effective parameters $J_1$ and $J_3$ show a strong asymmetry, which is
not present in the full set of interaction parameters shown in
Table~\ref{TAB:JsALK}.  This is mainly due to integrating out of the
fairly large coupling parameter $J_6$ in the model, which couples Cu
monomer with Cu dimer atoms and in this way provides an effective
asymmetry of $J_1$ and $J_3$ (inclusion of $J_6$ in addition to the
minimal model again results in nearly identical $J_1$ and $J_3$
values).

\begin{table}
\begin{tabular}{c||c|c|c|c}
$U$ [eV] & $J_1$ & $J_2$ & $J_3$ & $J_{\text{m}}$\\\hline
4 & 50.5 & 151.1 & 35.7 & 3.4 \\
6 & 30.0 & 85.4 & 21.0 & 3.0 \\
8 & 17.9 & 43.9 & 12.0 & 2.4 
\end{tabular}
\caption{{Effective exchange constants in K for a minimal model
including only $J_1$, $J_2$, $J_3$, and $J_{\text{m}}$ obtained via
statistical averaging (see text).}}
\label{TAB:JeffALK}
\end{table}

\subsection{DMRG calculations}

\label{sec:DMRG}

The theoretical magnetization curve in Fig.~3~(b) has been
obtained with the static density-matrix renormalization
group (DMRG) method~\cite{DMRGa,DMRGb} using
$m=300$ states per block and four sweeps in each magnetization sector.
The theory curves in Figs.~3~(c),
3~(d), and S1 have been obtained using
transfer-matrix DMRG (TMRG~\cite{TMRG96,TMRG}) for the infinite system and
$m=300$. Note that these are in agreement with previous TMRG
computations~\cite{GuSu1,GuSu2} for the parameters of
Ref.~\cite{KikuchiA}.

The transverse dynamic structure factor of
Fig.~3~(e) has been computed by dynamic
DMRG~\cite{dDMRG} for the parameters of Table~I, line {\bf
  3}, using open chains with $N=60$ sites, up to $m=200$ states
per block and two sweeps per energy point.
Note that for a meaningful comparison of scattering intensities with
experiment, we had to take the precise positions of the copper atoms
in azurite into account and use the same momentum perpendicular to the
chain direction as in the experiment~\cite{azuritINS}.

\subsection{Experiment}

The specific heat of a plate-like azurite single crystal with the total mass
of $0.36$~mg was measured, employing an ac-calorimetry according
to Ref.~\cite{Sullivan68}. The data were
taken in the temperature range $1.6~\text{K} \le T \le 30$~K and in magnetic
fields up to $8$~T. The experiments were performed using a home-built AC-calorimeter
especially designed for small plate-like samples. The sample holder,
consisting of a resistive thermometer (Cernox CX-1080-BG) and a heater, is
attached to a $^4$He-bath cryostat equipped with a superconducting magnet.

The magnetic susceptibility of azurite was measured in the temperature range
between $2~\text{K} \le T \le 300$~K and in magnetic fields up to $H = 4$~T
using a
Quantum Design SQUID magnetometer. The orientation of the single crystal
(mass $55.26$~mg) with respect to the external field was $H \perp b$-axis. The data
were corrected for the temperature-independent diamagnetic core
contribution, according to Ref.~\cite{Kahn93} and the magnetic
contribution of the sample holder. The latter was determined from an
independent measurement.

\subsection{Perturbative treatment of interchain coupling}

\label{sec:pert}

For a large and antiferromagnetic $J_2$, one can use perturbative
arguments to integrate out the copper dimers and generate effective
interactions between the monomer copper atoms. In the limit of
infinite $J_2$, the two spins on the corresponding dimer bond are in
their singlet state $\frac{1}{\sqrt{2}}\,\left( |\uparrow\downarrow
  \rangle - |\downarrow\uparrow \rangle \right)$.  In this limit, the
only interaction between the monomer spins is $J_{\text{m}}$. However,
one can use degenerate perturbation theory in $J_i/J_2$ to generate
further interactions between the monomer spins.

The second-order contribution to the monomer-monomer interactions
within a chain is known~\cite{HoL} to be given by
\begin{equation}
\tilde{J}_{\text{m}} = \frac{(J_1-J_3)^2}{2\,J_2}\, .
\label{eq:JmEff}
\end{equation}
This effective interaction enhances the bare interaction
$J_{\text{m}}$ between the monomers along the chain provided that $J_1
\ne J_3$.

The interactions $J_4$ and $J_7$ connect dimers of neighboring chains
(see Fig.~1~(d)). Accordingly, they contribute
to interchain monomer-monomer exchange only in third order in
perturbation theory and generate exchanges $\propto J_1^2\,J_4/J_2^2$,
$J_1\,J_3\,J_4/J_2^2$, $J_1^2\,J_7/J_2^2$ and
$J_1\,J_3\,J_7/J_2^2$. Using the values of the $J_i$ in
Table~I, line {\bf 1}, these effective interchain exchanges
are estimated to be at most on the order of $0.3~\text{K} \approx
J_{\text{m}}/10$ and thus can be neglected safely.

By contrast, $J_5$ and $J_6$ contribute in second order perturbation
theory to interchain coupling since they connect dimers with monomers
of the neighboring chains (see Fig.~1~(c)).
The contribution from $J_5$ is given by $(J_1+J_3)\,J_5/(2\,J_2)$.
Inserting the numbers from Table~I, line {\bf 1}, this
again turns out to be on the order of $0.2~\text{K} \approx
J_{\text{m}}/10$, {\it i.e.},
also $J_5$ is sufficiently small to be neglected safely.

The exchange constant $J_6$ also contributes terms proportional to
$J_1\,J_6/J_2$ and $J_3\,J_6/J_2$ to interchain effective
monomer-monomer coupling. Inserting again the values of $J_1$, $J_2$, $J_3$,
and $J_6$ in
Table~I, line {\bf 1} into the second-order expression, we
now obtain a contribution on the order of $1.4~\text{K} \approx
J_{\text{m}}/2$ to the effective
interchain monomer-monomer coupling. On the one hand, this is still
sufficiently small not to give rise to relevant dispersion of the
excitations perpendicular to the chains, in agreement with inelastic
neutron scattering on azurite~\cite{azuritINS}. On the other hand,
this value is too large to neglect $J_6$ completely.

In fact, in a mean-field picture, the monomer moments influence
the effective monomer-monomer exchange along the neighboring chains.
The reason is that the interchain couplings connect the monomer spins
only to one of the dimer spins on the neighboring chains, thus
breaking the symmetry of the exchange process along the chains
and giving rise to corrections to (\ref{eq:JmEff}). In this mean-field
picture, the interchain coupling $J_6$ has the same effect as the
intrachain coupling $J_1$.

These arguments suggest that one may go from the full
three-dimensional model to an effective chain model by neglecting
$J_4$, $J_5$, and $J_7$, and adding $J_6$ to $J_1$. The difference
between lines {\bf 1} and {\bf 2} of Table~I or
Tables~\ref{TAB:JsALK} and \ref{TAB:JeffALK} can indeed be understood
at least qualitatively
in this way although the reduction has been performed in a completely
different manner.

\subsection{Specific heat}

\label{sec:specificHeat}

\begin{figure}[tb!]
\begin{center}
\includegraphics[width=\columnwidth]{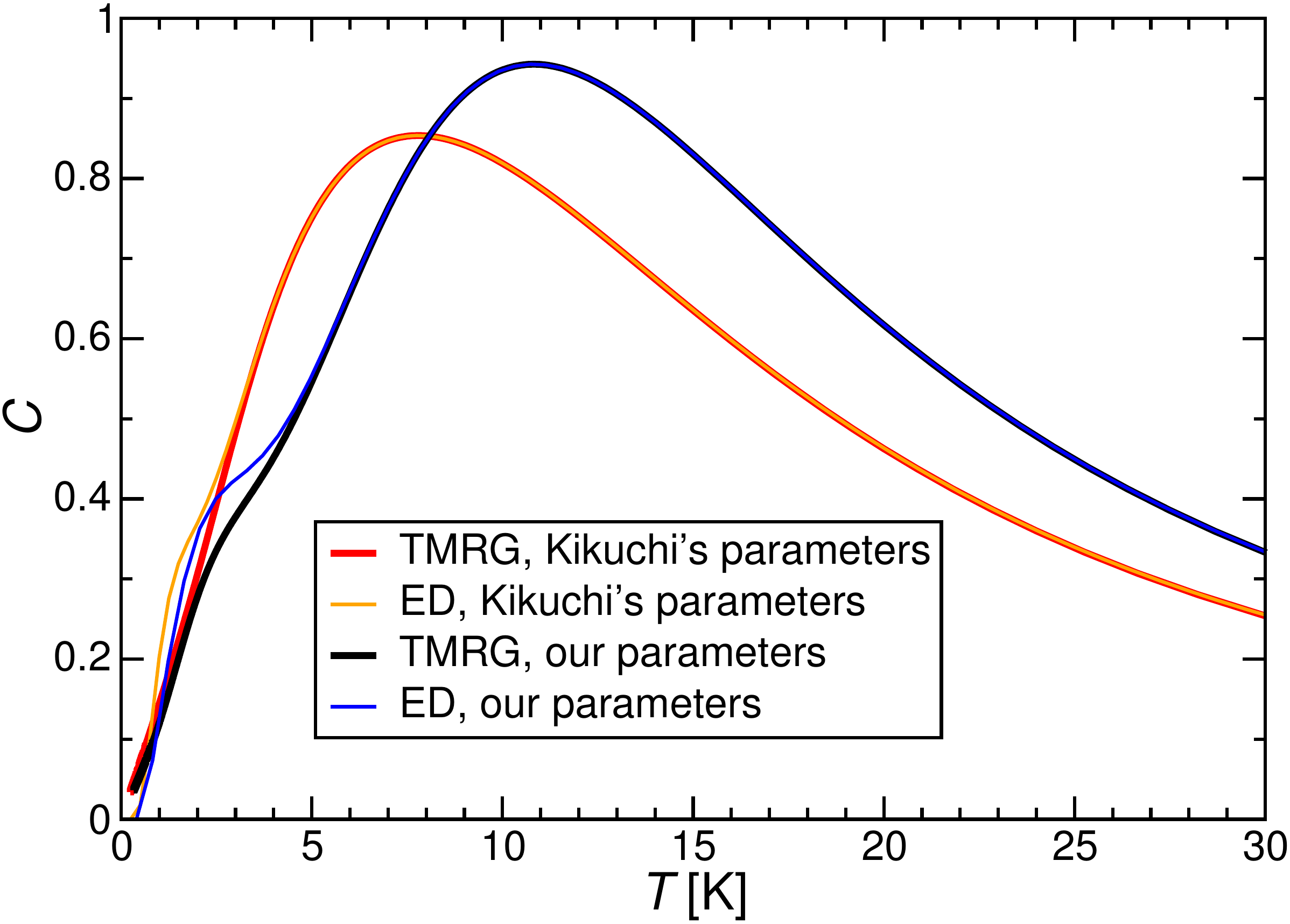}
\end{center}
\caption{{
    TMRG and ED  results for the zero-field specific heat per spin. The
    ED computations were performed for $N=18$ spins.  We show results
    both for our new parameter set, Table~I, line {\bf 3},
as well as for Kikuchi's original parameter set Table~I, line {\bf 4}
$J_1= 19$~K, $J_2=24$~K, $J_3=8.6$~K, and $J_{\text{m}}=0$~\cite{KikuchiA}.
}}
\label{fig5}
\end{figure}

Experimentally, two anomalies have been observed in the magnetic
specific heat at $T\approx 18$~K~\cite{KikuchiA} and
$T\approx 4$~K~\cite{azuritINS,KikuchiA} (compare also top panel of
Fig.~3~(d)).
Fig.~\ref{fig5} shows TMRG results for the specific heat per spin $C$
in zero magnetic field. For our new parameter set,
Table~I, line {\bf 3} (black line in Fig.~\ref{fig5}),
we find a maximum of $C$ at a temperature slightly above $10$~K and
a low-temperature feature at $T \approx 3$~K.
Although this does
not reproduce the experimental temperatures exactly, it is
in better agreement with the experimental findings than the results for the
original parameter set of Ref.~\cite{KikuchiA} (red line in
Fig.~\ref{fig5}).

Fig.~\ref{fig5} includes exact diagonalization (ED) results for
rings with $N=18$ spins. We observe that finite-size effects have no
visible effect for $T \gtrsim 6$~K.

\subsection{Structure of the 1/3 plateau}

\label{sec:strucM1o3}

\begin{table}[tb!]
\begin{center}
\begin{tabular}{c||c|c}
$N$ &
monomer $\langle S^z_i \rangle$ &
dimer $\langle S^z_i \rangle$  \\ \hline
$18$ & $0.47342867$ & $0.01328567$ \\
$24$ & $0.47343148$ & $0.01328426$ \\
$30$ & $0.47343154$ & $0.01328423$ \\
$36$ & $0.47343154$ & $0.01328423$ \\
\end{tabular}
\end{center}
\caption{{Structure of the $M=1/3$ plateau state for rings with
    $N$ sites and the parameters
    in  line {\bf 3} of Table~I.
\label{tab:M1o3}
}}
\end{table}

The 1/3 plateau state of azurite has been characterized using
NMR~\cite{azuritNMR}, which amounts to a measurement of the
expectation values $\langle S^z_i \rangle$. This NMR study showed that
the dimer spins are essentially in their singlet state with just
10{\%} spin polarization on the dimers. Correspondingly, the monomer
spins are almost polarized on the 1/3 plateau.

Using ED for rings with $N=18$, $24$, $30$, and $36$ sites and our
parameters line {\bf 3} of Table~I, we find the
structure of the $M=1/3$ plateau state presented in Table~\ref{tab:M1o3}.
We observe that the numerical results for the expectation values converge
rapidly with system size and read off that the dimer spins are about
2.7{\%} polarized each.  This is only slightly smaller than the 10{\%}
observed in Ref.~\cite{azuritNMR}.  We note that the NMR
experiment~\cite{azuritNMR} involved a rotation around the
crystallographic $a$-axis
and speculate that this gives rise to non-commuting fields which
enhance the dimer polarization as compared to the ideal Heisenberg
model.

\subsection{Excitation spectrum on the 1/3 plateau}

\begin{figure}[tb!]
\begin{center}
\includegraphics[width=\columnwidth]{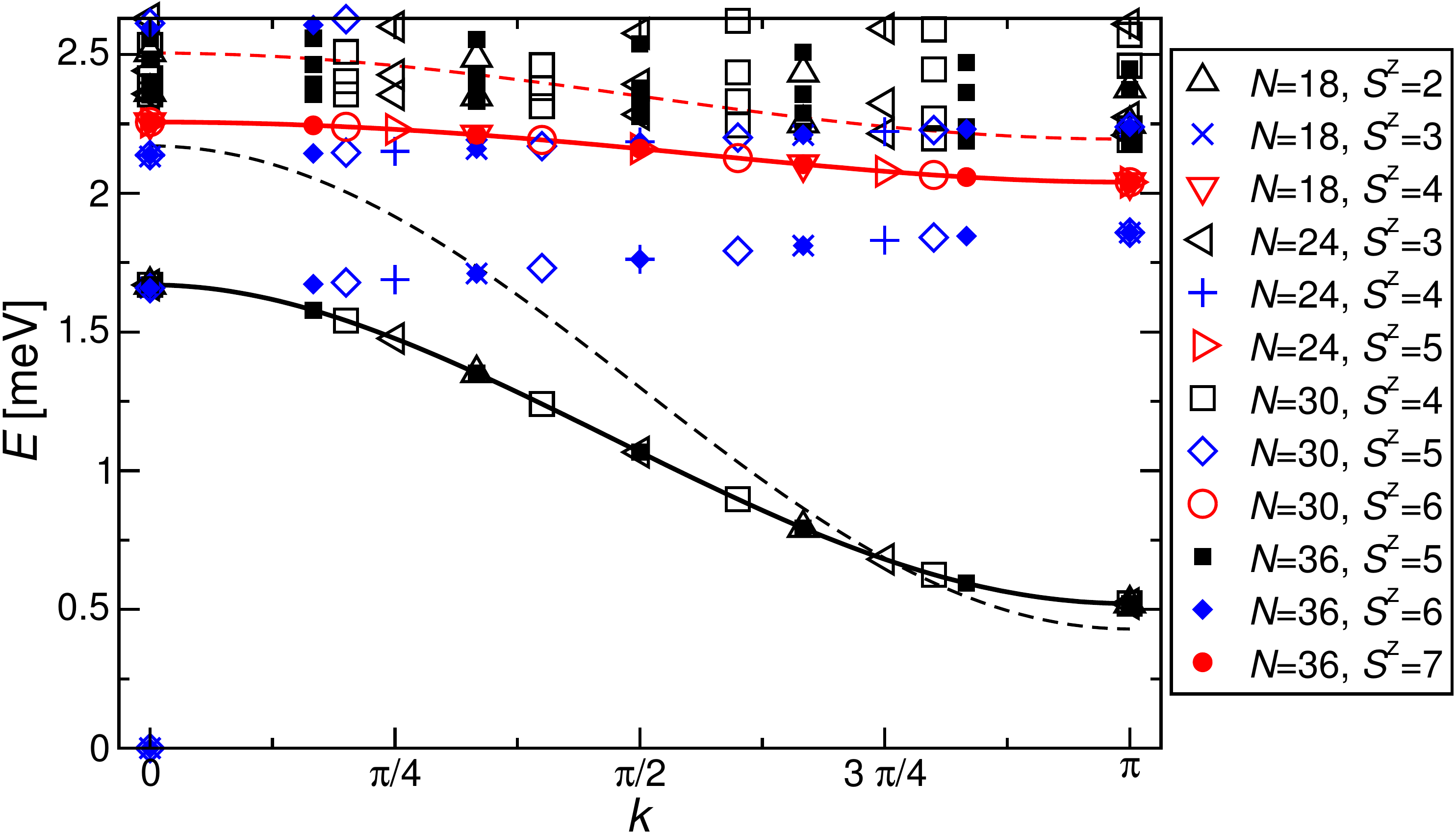}
\end{center}
\caption{{
    Excitation spectrum at $H=14$~T 
    as computed by exact diagonalization with the parameters in
    Table~I, line {\bf 3}.
Solid black (red) lines connect the lowest excitations with spin quantum
numbers smaller (larger) by one than that of the 1/3 plateau state.
Dashed lines indicate the location of the experimental result~\cite{azuritINS}
for the corresponding excitations.
}}
\label{fig3}
\end{figure}

\begin{figure}[tb!]
\begin{center}
\includegraphics[width=\columnwidth]{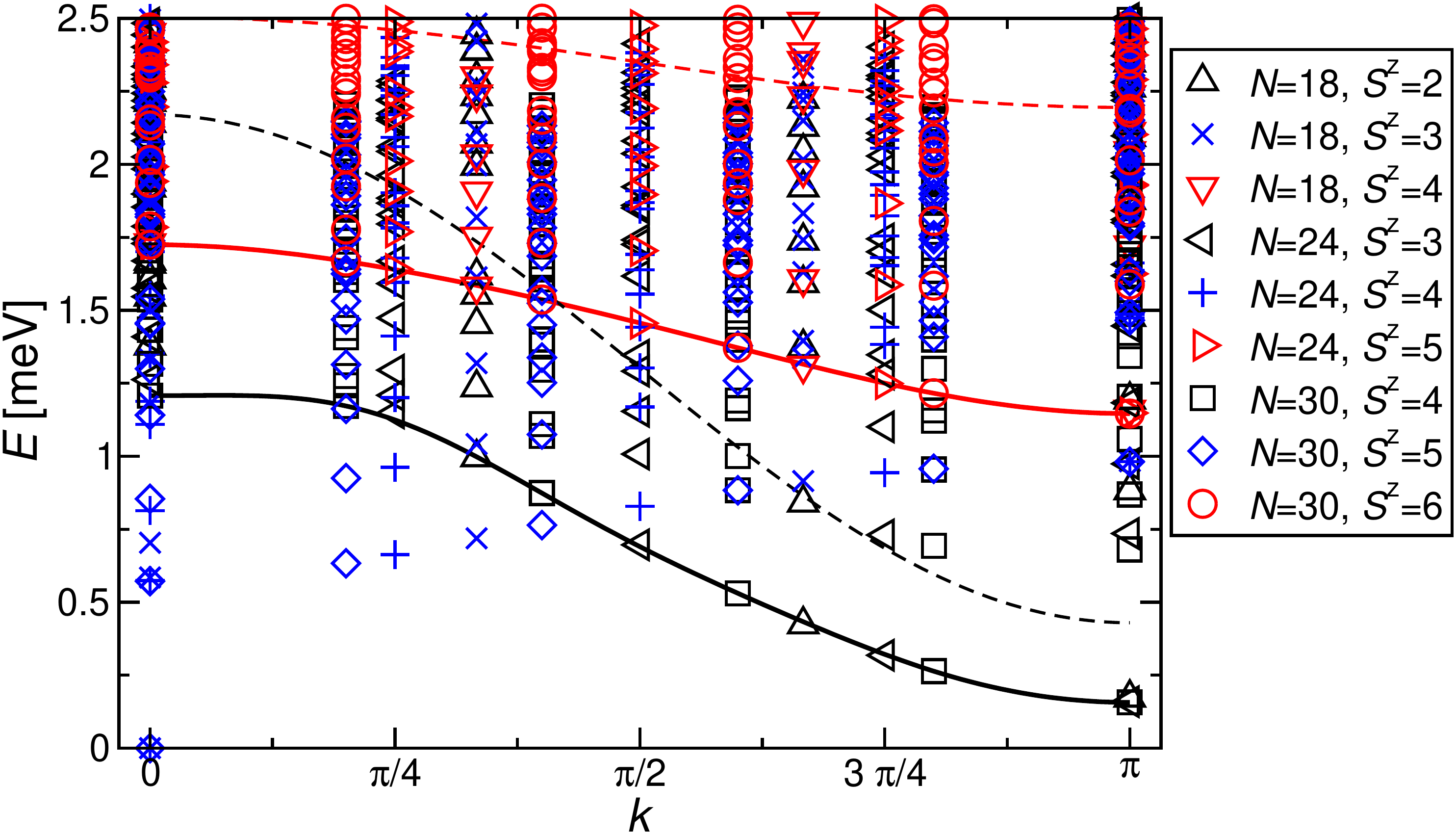}
\end{center}
\caption{{
    Same as Fig.~\ref{fig3}, but
    for Kikuchi's original parameter set
    Table~I, line {\bf 4}, $J_1= 19$~K, $J_2=24$~K,
    $J_3=8.6$~K, and $J_{\text{m}}=0$~\cite{KikuchiA}.  }}
\label{fig3kikuchi}
\end{figure}

\begin{figure}[tb!]
\begin{center}
\includegraphics[width=\columnwidth]{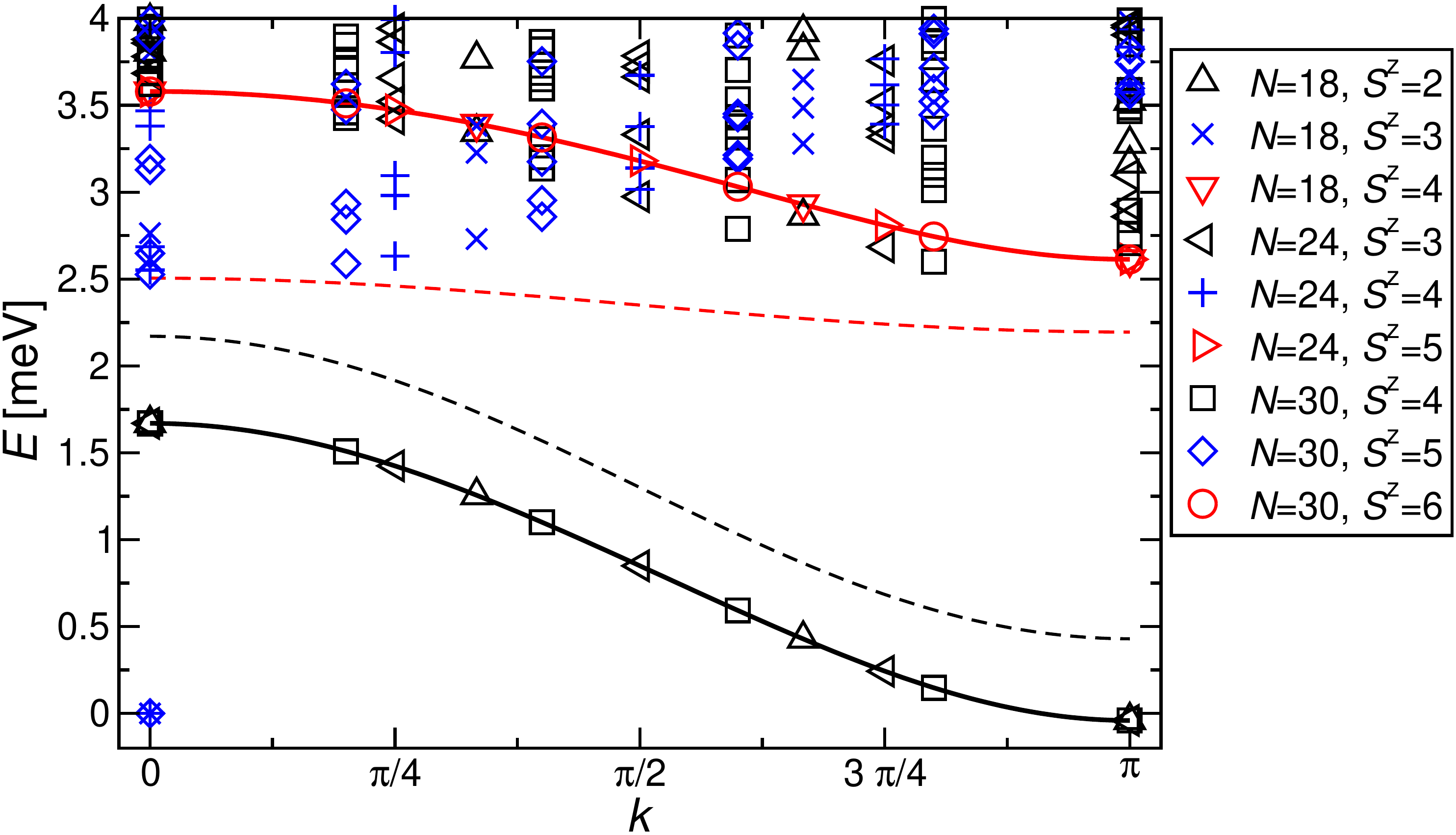}
\end{center}
\caption{{
    Same as Fig.~\ref{fig3}, but
    for the parameter set of Gu and
    Su~\cite{GuSu1,GuSu2}: $J_1= 23$~K, $J_2=43.7$~K, $J_3=-9.3$~K,
    and $J_{\text{m}}=0$.  }}
\label{fig3GuSu}
\end{figure}

\begin{figure}[tb!]
\begin{center}
\includegraphics[width=\columnwidth]{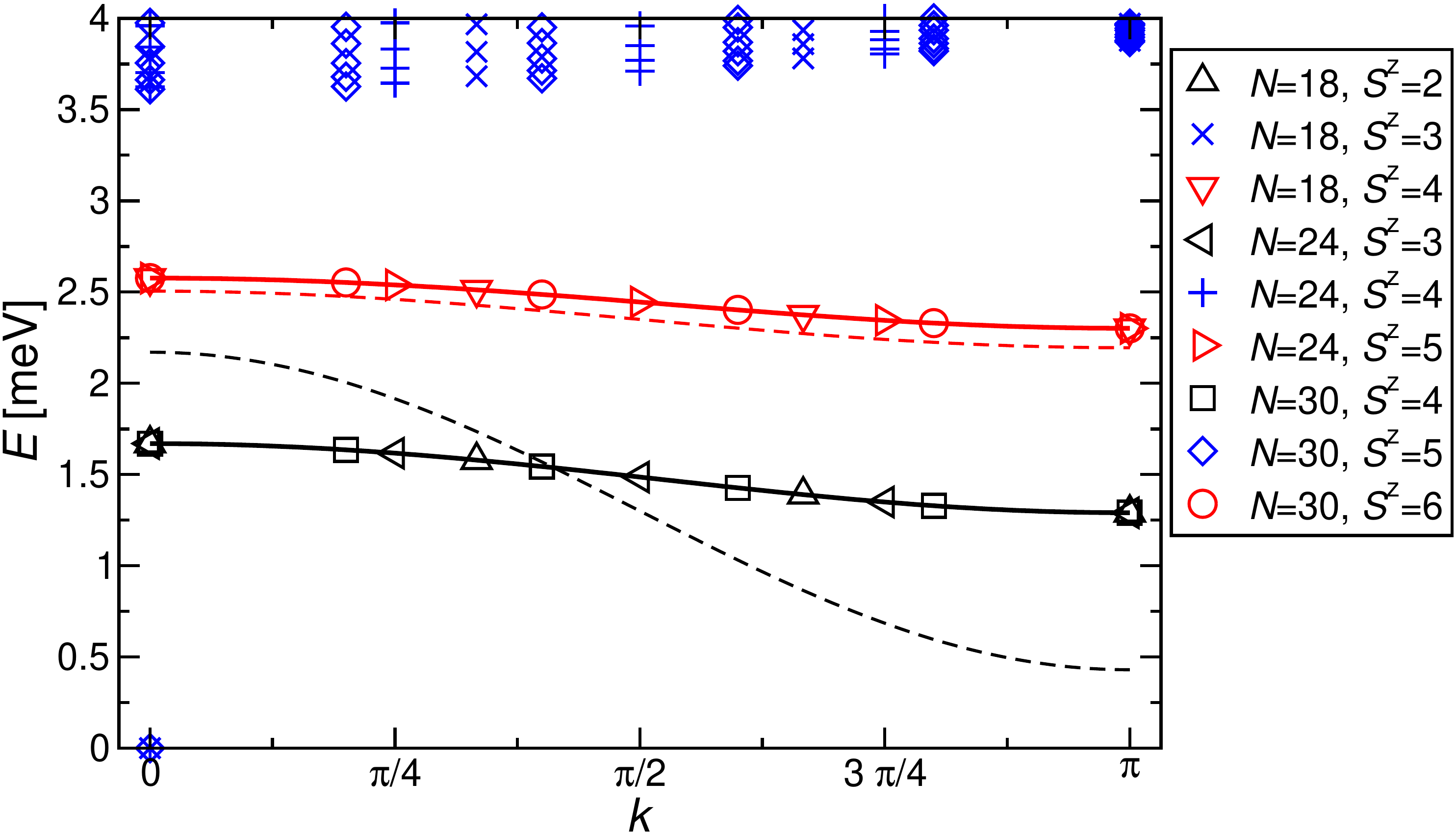}
\end{center}
\caption{{
    Same as Fig.~\ref{fig3}, but
    for a first parameter set proposed in
    Ref.~\cite{azuritINS}: $J_1= 1$~K, $J_2=55$~K, $J_3=-20$~K, and
    $J_{\text{m}}=0$.  }}
\label{fig3rule1}
\end{figure}

\begin{figure}[tb!]
\begin{center}
\includegraphics[width=\columnwidth]{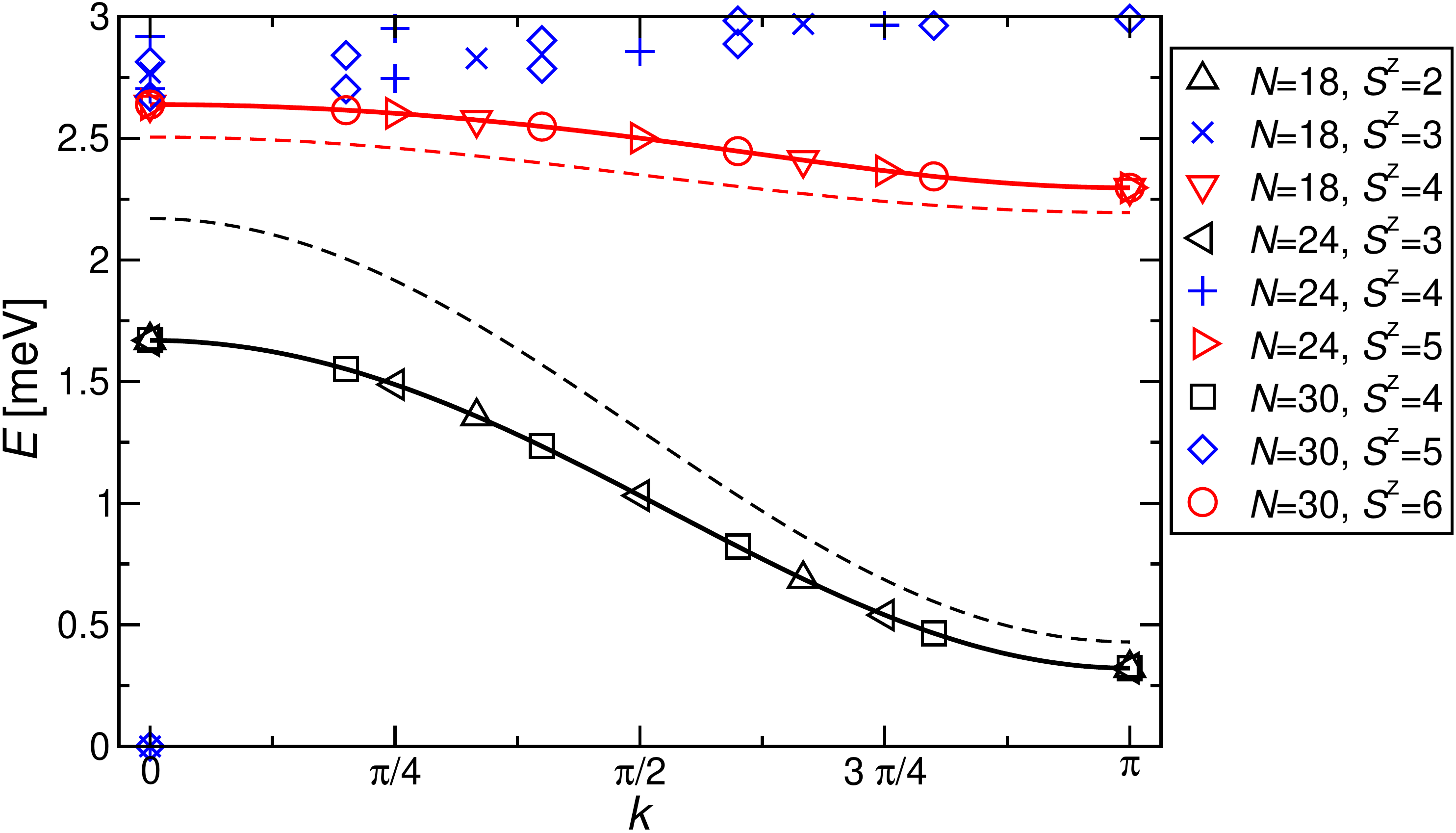}
\end{center}
\caption{{
    Same as Fig.~\ref{fig3}, but
    for the second parameter set proposed in
    Ref.~\cite{azuritINS}: $J_1= 1$~K, $J_2=55$~K, $J_3=-20$~K, and
    $J_{\text{m}}=6.5$~K.  }}
\label{fig3rule2}
\end{figure}

The excitation spectrum above the 1/3 plateau of azurite has been
probed by inelastic neutron scattering at
$H=14$~T~\cite{azuritINS}. These experiments observed two cosine-like
bands of magnetic excitations with a minimum energy at the
antiferromagnetic wave vector $k=\pi$. The two bands are sketched
by the dashed lines in Figs.~\ref{fig3}--\ref{fig3rule2}. They are
centered
around $\approx 1.3$~meV and $\approx 2.35$~meV and have a width of
$20.2$~K and $3.6$~K, respectively. In particular the ratio of the
bandwidths of the upper and lower bands is $1.8/10.1 \approx 1/5.6$.

Figs.~\ref{fig3}--\ref{fig3rule2} show the excitation spectrum as a
function of momentum $k$ along the chain direction on the 1/3 plateau
computed by exact diagonalization with periodic boundary conditions.
Black, blue, and red symbols correspond to excitations with $\Delta
S^z=-1$, $0$, and $1$, respectively. The blue symbol at $k=0$
and energy $E=0$
corresponds to the ground state of the 1/3 plateau. A Fourier
analysis of the lowest $\Delta S^z= \pm 1$ excitations for $N=30$
sites yields the solid lines in Figs.~\ref{fig3}--\ref{fig3rule2}.
One observes that the lowest $\Delta S^z= \pm 1$ excitations
collapse onto these lines for all sizes $N$, demonstrating that
the main effect of a finite system size $N$ on these excitations
is a discretization of the allowed values of the momentum $k$
(see also~\cite{MiLu}).

First let us look at our final parameter set Table~I, line
{\bf 3}.  The lowest black and red excitation in Fig.~\ref{fig3}, {\it i.e.},
the two solid curves
correspond to the two dispersion curves already observed in
Fig.~3~(e).  Note that energy and momentum
resolution in Fig.~3~(e) is essentially limited
by the open ends of the finite-size chains which were used for the
dynamic DMRG computations. Evidently, inspection of the
momentum-resolved bare energy levels shown in Fig.~\ref{fig3} yields
better energy-momentum resolution at the expense of losing information
about the neutron intensities of the excitations.  Indeed, in
Fig.~\ref{fig3} we see a large number of excitations at energies above
$2$~meV and only the dynamic structure factor of
Fig.~3~(e) shows that they have very little
contribution to the inelastic neutron cross section.  {}From the bare
energy levels of Fig.~\ref{fig3}, we read off a bandwidth ratio of
$1/5.3$ which is very close to the experimental
value~\cite{azuritINS}.

Fig.~\ref{fig3kikuchi} shows the expected excitation spectrum for
Kikuchi's original parameter set Table~I, line {\bf
  4}~\cite{KikuchiA}. Note that we have computed only the 20 lowest
excitations in some sectors for $N \ge 24$ and that the density of
states is already quite large at energies above $1$~meV in the present
case.  Hence, some levels may be missing in Fig.~\ref{fig3kikuchi} at
energies $E > 2$~meV for $S^z \le N/6$ and $N \ge 24$. We nevertheless
keep this region in order to be able to show the location of the
excitations observed by inelastic neutron scattering on
azurite~\cite{azuritINS} (dashed lines).  In the present case it is
not so easy to distinguish two cosine-like bands in the numerical
results.  If one uses the lowest black and red energy level,
respectively, one finds a bandwidth ratio close to $1/1.9$, quite far
off the experimental result~\cite{azuritINS}.

Further proposals of parameters sets~\cite{azuritINS,GuSu1,GuSu2}
contain a ferromagnetic $J_3$.  Ref.~\cite{Whangbo} already pointed
out that a ferromagnetic $J_3$ is hard to reconcile with the crystal
structure of azurite given the $d_{x^2-y^2}$ character of the relevant
copper orbitals. We will nevertheless look at the excitation spectra
for these parameter sets and demonstrate that they are either
inconsistent or at least yield less good agreement with the neutron
scattering experiments~\cite{azuritINS} than our final parameter set
given in Table~I, line {\bf 3}.

The parameter set of Gu and Su~\cite{GuSu1,GuSu2} $J_1=23$~K,
$J_2=43.7$~K, $J_3^{\text{xy}}=-6.9$~K, $J_3^{\text{z}}=-11.73$~K,
$J_{\text{m}}=0$ has an artificially large magnetic anisotropy in the
supposedly ferromagnetic $J_3$. Replacing this by an average value
$J_3=-9.3$~K, we find the excitation spectrum shown in
Fig.~\ref{fig3GuSu}.  Not only is the bandwidth ratio of approximately
$1/1.8$ again far away from the experimental result~\cite{azuritINS},
but in this case the upper excitation branch is about $1$~meV ($\approx
10$~K) too high in energy.

Finally, Ref.~\cite{azuritINS} tried to invert perturbative results
for the effective monomer-monomer and dimer-dimer exchanges along the
chain in order to propose $J_1=1$~K, $J_2=55$~K and a ferromagnetic
$J_3=-20$~K.  We discuss two variants of these parameters, starting in
Fig.~\ref{fig3rule1} with $J_{\text{m}}=0$. In this case, we find a
bandwidth ratio $1/1.4$ which is clearly inconsistent with the
experimental result.  However, it was already proposed in
Ref.~\cite{azuritINS} to improve this behavior by adding a
$J_{\text{m}}=6.5$~K. The result with such a $J_{\text{m}}$ included
is shown in Fig.~\ref{fig3rule2}. While inclusion of
$J_{\text{m}}=6.5$~K improves the agreement with
experiment~\cite{azuritINS}, the result is not quite as good for our
final parameter set Table~I, line {\bf 3}. In particular,
the bandwidth ratio is just $1/4$.

To summarize the discussion of this subsection, we have demonstrated
that our final parameter set Table~I, line {\bf 3} yields
the best agreement with inelastic neutron scattering on the 1/3
plateau~\cite{azuritINS} among the proposals
of Refs.~\cite{azuritINS,KikuchiA,GuSu1,GuSu2}.  In particular, inelastic
neutron scattering is  inconsistent with the parameters
proposed in Refs.~\cite{KikuchiA,GuSu1,GuSu2}.

\subsection{Perspectives}

\label{sec:perspectives}

There are some
further refinements of the model for azurite to be implemented in future
investigations. Firstly, we have argued interchain
coupling to be unimportant for a basic description of azurite, but, although
small, it is present and likely to be responsible for the following
features: (i) The strong curvature at the lower edge of 1/3 plateau in
the theoretical magnetization curve of Fig.~3~(b)
versus the smoother behavior observed in the experiment for $H
\perp b$ is characteristic for one- versus higher-dimensional
physics~\cite{DzNe,PoTa,BaBra}.  (ii) An ordering transition at
temperatures slightly below $2$~K~\cite{KikuchiA,Forstat,gibson10,rule10}
is evident in Fig.~3~(d), {\it i.e.},
interchain coupling affects thermodynamic properties at temperatures
of a few Kelvin. In view of the success of the effective one-dimensional
model Table~I, line {\bf 3}, we are confident that our
prediction of Table~I, line {\bf 1} for the exchange ratios
of the full three-dimensional model
is also reliable. Indeed, it would be
very interesting to compare the predictions of this three-dimensional model
for the zero-field ordered state with the corresponding recent experimental
observations \cite{rule10}. However, this will require very different
methods from the present work and thus is an interesting topic for
future investigations.

Secondly, we have neglected magnetic anisotropy in
the theoretical model although experiments~\cite{KikuchiA} show that
it is present in azurite and affects magnetic properties for a
magnetic field parallel to the crystallographic $b$-axis at an energy
scale of a few Kelvin.

Finally, we would like to emphasize that we find all $J_i$
antiferromagnetic with similar values of $J_1$ and $J_3$, thus placing
azurite in a highly frustrated parameter regime. This is reflected by
the almost localized nature of the dimer excitations.  These
excitations will become low-energy excitations in magnetic fields
around $32$~T, and one expects related unusual thermodynamic behavior
like an enhanced magnetocaloric effect~\cite{LTP2007,Amcal11}. This calls for
additional thermodynamic measurements close to the saturation field
of azurite.


\begin{thebibliography}{99}
\bibitem{comment} The orbital assignment
is given in the local reference frame of Cu
where the $ab$ plane corresponds to the CuO$_4$ plane with $a$
parallel to a Cu-O bond.
\bibitem{footnote} We would like to emphasize that we have made no further
adjustments to the magnetic field $H=14$~T although the critical field at the
lower edge of the $M=1/3$ plateau is reproduced less accurately
than other features (compare Fig.~\ref{fig:comparison}~(b)). Therefore, we
attach more importance to the bandwidth ratio than the absolute position of
the excitation bands.
\end{thebibliography}
\end{document}